\documentclass[twocolumn]{elsart}
\usepackage[dvips]{graphicx}

\begin{document}
\def\Journal#1#2#3#4{{#1} {\bf #2}, #3 (#4)}

\def\NCA{\em Nuovo Cimento}
\def\NIM{\em Nucl. Instrum. Methods}
\def\NIMA{{\em Nucl. Instrum. Methods} A}
\def\NPB{{\em Nucl. Phys.} B}
\def\PLB{{\em Phys. Lett.}  B}
\def\PRL{\em Phys. Rev. Lett.}
\def\PRD{{\em Phys. Rev.} D}
\def\ZPC{{\em Z. Phys.} C}
\def\APJ{\em Astrophys. J.}
\def\SNP{\em Sov. Jour. Nucl. Phys.}
\def\st{\scriptstyle}
\def\sst{\scriptscriptstyle}
\def\mco{\multicolumn}
\def\epp{\epsilon^{\prime}}
\def\vep{\varepsilon}
\def\ra{\rightarrow}
\def\ppg{\pi^+\pi^-\gamma}
\def\vp{{\bf p}}
\def\ko{K^0}
\def\kb{\bar{K^0}}
\def\al{\alpha}
\def\ab{\bar{\alpha}}
\def\be{\begin{equation}}
\def\ee{\end{equation}}
\def\bea{\begin{eqnarray}}
\def\eea{\end{eqnarray}}
\def\CPbar{\hbox{{\rm CP}\hskip-1.80em{/}}}

\newcommand{\plumin}[2]{^{+#1}_{-#2}}
\newcommand{\plumi}[2]{\matrix{+#1\\[-3pt]-#2}}
\newcommand{\stth}{\sin^2 2\theta}
\newcommand{\cth}{\cos^2 \theta}
\newcommand{\sth}{\sin^2 \theta}
\newcommand{\csth}{\cos \theta\sin \theta}
\newcommand{\dm}{\Delta m^2}
\newcommand{\tsun}{\theta_{\mbox{\tiny Sun}}}
\newcommand{\csun}{\cos \tsun}
\newcommand{\phiosc}{\phi^{\mbox{\tiny osc}}_{i,z}}
\newcommand{\phissm}{\phi^{\mbox{\tiny SSM}}_i}
\newcommand{\delcor}{\delta_{\mbox{\tiny corr}}}

\begin{frontmatter}
\newcounter{foots}
\newcommand{\newad}{\addtocounter{foots}{1}$^{,\protect\fnsymbol{foots}}$}
\newcommand{\oldad}{$^{,\protect\fnsymbol{foots}}$}
\title{Determination of Solar Neutrino Oscillation Parameters using
1496 Days of Super-Kamiokande-I Data}

\author[icrr]{S.~Fukuda},     \author[icrr]{Y.~Fukuda},
\author[icrr]{M.~Ishitsuka},  \author[icrr]{Y.~Itow},
\author[icrr]{T.~Kajita},     \author[icrr]{J.~Kameda},
\author[icrr]{K.~Kaneyuki},   \author[icrr]{K.~Kobayashi}, 
\author[icrr]{Y.~Koshio},     \author[icrr]{M.~Miura},
\author[icrr]{S.~Moriyama},   \author[icrr]{M.~Nakahata}, 
\author[icrr]{S.~Nakayama},   \author[icrr]{T.~Namba},
\author[icrr]{A.~Okada},
\author[icrr]{N.~Sakurai},    \author[icrr]{M.~Shiozawa},
\author[icrr]{Y.~Suzuki},     \author[icrr]{H.~Takeuchi}, 
\author[icrr]{Y.~Takeuchi},   \author[icrr]{Y.~Totsuka},
\author[icrr]{S.~Yamada},
\author[bu]{S.~Desai},        \author[bu]{M.~Earl},
\author[bu]{E.~Kearns},       \author[bu]{M.D.~Messier}\newad,
% ...
\clearpage
\address[icrr]{Institute for Cosmic Ray Research, University of Tokyo,
               Kashiwa, Chiba 277-8582, Japan}
\address[bu]{Department of Physics, Boston University,
             Boston, MA 02215, USA}
% ...

\begin{abstract}
A number of different fits to solar neutrino mixing and mass square
difference were performed using
1496 days of Super-Kamiokande-I's solar neutrino data.
These data select
two allowed areas at large neutrino mixing
when combined with
either the solar $^8$B flux prediction of the standard solar
model or the SNO interaction rate measurements.
A global fit combining 
SK data with the solar neutrino interaction rates
measured by Homestake, SNO, Gallex/GNO and SAGE 
prefers a single allowed area, the Large Mixing Angle
solution, at the 98.9\% confidence level. The mass square
difference
$\Delta m^2$ between the two mass eigenstates ranges from about
$3$ to $19\times10^{-5}$eV$^2$, while the mixing angle
$\theta$ is in the range of $\tan^2\theta\approx$0.25--0.65.
\end{abstract}
\end{frontmatter}

\section{Solar Neutrino Oscillations}
While a number of previous solar neutrino
experiments~\cite{cl,kam,gallex,sage}
have measured a smaller neutrino interaction rate than that
predicted by the Standard Solar Model (SSM) \cite{ssm}, in 1996
Super-Kamiokande (SK) began collecting data of unprecedented precision
and quantity. In addition to a reduced interaction rate of
solar $^8$B neutrinos (originating from $\beta^+$ decay
of $^8$B nuclei)
with electrons in the detector,
SK measured with high precision the spectrum
of the recoiling electrons as well as time variations of this
reduced rate~\cite{sk}.

\begin{figure*}[hbt]
\includegraphics[height=7.8cm,clip]{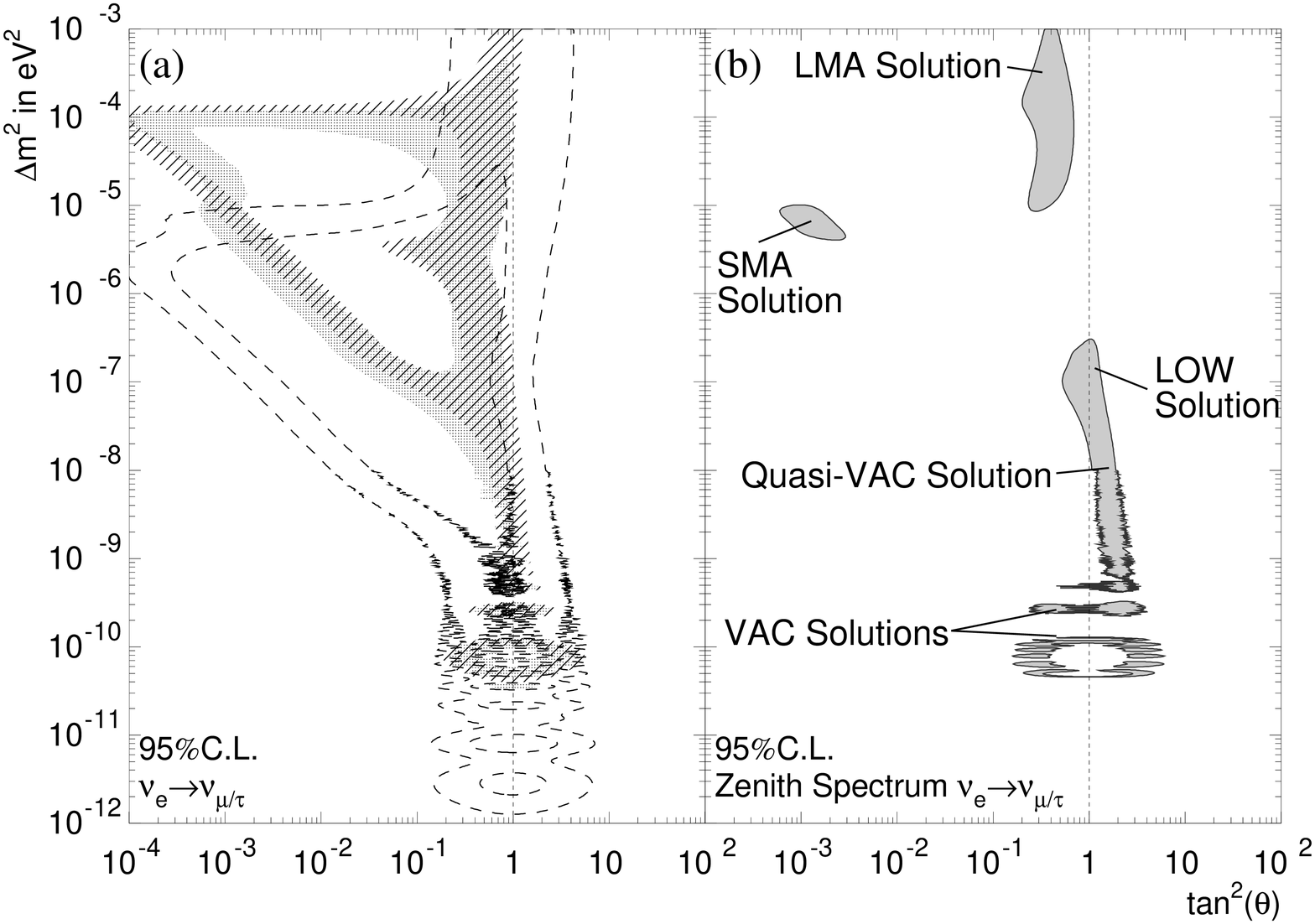}

\caption{
(a) Allowed areas from other experiments
measuring the solar $\nu_e$ flux.
The shaded area uses only Homestake data, while the 
hatched area uses only the SNO charged-current rate.
Overlaid (inside dashed
lines) is the region allowed by
Gallex/GNO and SAGE.
(b) Allowed regions
from a combined fit to these charged-current rates.
All contours in this and other figures are 95\% C.L.
\label{fig:singleexp}
}
\end{figure*}

Neutrino flavor oscillations are able to explain all observed
solar neutrino interaction rates, including the rate
difference between SK and SNO~\cite{sno}.
The large neutrino flavor mixing
between the second and third generation inferred from atmospheric
neutrino data~\cite{skatm} in conjunction with the absence of an oscillation
signal in the CHOOZ reactor neutrino experiment~\cite{chooz}
(which is sensitive to oscillations in a similar region of 
mass square difference)
requires the neutrino flavor mixing between the first and third
generation to be small. Solar neutrino oscillations can therefore
be approximated by a two-neutrino description with the parameters
$\theta$ (mixing angle) and $\Delta m^2$ (difference in mass square).
The mass eigenstate $m_1$ ($m_2$) is chosen to have the smaller
(bigger) mass \footnote{Alternatively,
it is possible to call the mass eigenstate $m_1$ ($m_2$) which has the larger
coupling to electrons (muons). In that case, $\theta$ ranges between
0 and $\pi/4$ and $\Delta m^2$ might be negative. The oscillation
probability remains invariant under the transformation 
$\theta^\prime=\frac{\pi}{2}-\theta$ and $\Delta m^{2\prime}=-\Delta m^2$.}.
Therefore, $\Delta m^2=m_2^2-m_1^2$ is always positive and
the mixing angle ranges between 0 and $\pi/2$.
For $\Delta m^2$ between $10^{-8}$eV$^2$ and
$10^{-3}$eV$^2$, matter densities in the sun and the earth can
strongly affect the oscillation probability. 
Consequently, the symmetry of the vacuum oscillation
probability around $\theta=\pi/4$ is broken.
For $\theta<\pi/4$,
resonant enhancement of the oscillations (MSW effect~\cite{msw})
can occur in the sun, while for $\theta>\pi/4$,
an anti-resonance can suppress the oscillations.
Since matter-dominated solutions usually don't
occur for $\theta>\pi/4$, we have not shown this
``dark side'' of the parameter space in previous reports.
Below $10^{-9}$eV$^2$,
the oscillation probability
is more affected by the oscillation phase than by matter effects.
A logarithmic scale of the variable $\tan^2\theta$
illustrates the mirror symmetry of the vacuum region
(below $\approx4\cdot10^{-10}$eV$^2$)
at the maximal mixing line.
Figure~\ref{fig:singleexp} (a) shows the experimentally
allowed areas using various measurements.

In addition to the suppression of the $\nu_e$ rates,
neutrino oscillations also produce other 
neutrino flavors. Furthermore, they can induce distortions
of the neutrino spectrum and time variations of the
solar neutrino fluxes. In the MSW region, the time variations
arise from matter effects inside the earth (daily variations).
In the vacuum regions, the time variations are the consequence
of the change of the oscillation phase due to the yearly
variation of the oscillation baseline -- the distance between the
sun and the earth. 

\begin{figure*}[hbt]
\includegraphics[height=7.8cm,clip]{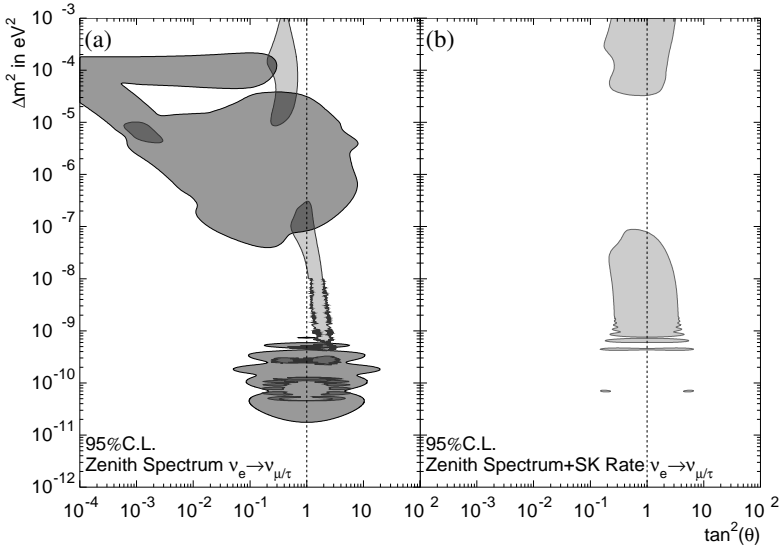}

\caption{
(a) SSM flux independent excluded areas (gray)
using the SK zenith spectrum shape alone
overlaid with the allowed regions of
Figure~\protect\ref{fig:singleexp} (b)
in light gray.
The overlap of both is shaded dark gray.
(b) Allowed areas using only SK data
and the SSM $^8$B neutrino flux prediction.
Both allowed regions indicate large neutrino mixing.
}
\label{fig:sk}
\end{figure*}

In this paper we report on the new SK data set of 1496 days,
which represents the completion of the analysis of all the data
from Super-Kamiokande-I. Previous SK energy spectrum and time-variation
measurements ruled out large regions of the oscillation parameter
space allowed by other experiments~\cite{osc}. When combined with
the SSM, these measurements also restricted $\tan^2\theta$ to be
bigger than $\approx$0.2 disfavored oscillation to a sterile neutrino.
Our final data set confirms our previous measurements and further
constrains the range of possible oscillation parameters.

Even stronger
constraints can be obtained by combining SK results with the
rate measurements of other experiments. In particular, the
combination of SNO's measurement of the solar $^8$B $\nu_e$
interaction rate~\cite{snonc} with the SK rate yields
additional information~\cite{sno,venice} about $^8$B flavor
conversion, since SK also has sensitivity to
other active neutrino flavors. If the $\nu_e$ survival probability
is moderately small for $^8$B neutrinos
(less than about 50\%) then SK can
observe a significant excess due to the presence of other active species
in its precise measurement of the elastic scattering rate.
A similar signature can be observed in the combination of
SK and the Homestake rate. In this case, however, the lower energy threshold
of Homestake and its resulting sensitivity to other than $^8$B
neutrinos lead to an ambiguity between spectral distortion
and appearance signatures and therefore to a larger uncertainty
in the observed excess.

\section{Parameter Estimation for Two-Neutrino Oscillations}
Without SK data, the strongest
constraint of the parameters is
a $\chi^2$ fit to the charged-current neutrino rates from
Gallex/GNO, SAGE (combined into a single ``Gallium'' rate),
Homestake (``Chlorine'') and SNO experiments~\cite{ap}.
The light gray areas of
Figure~\ref{fig:singleexp} (b) use only these three rates
and their SSM-based predictions.
The regions can be understood as the ``overlap''
of the three allowed regions of Figure~\ref{fig:singleexp} (a).
Above $\Delta m^2\approx 10^{-5}$eV$^2$ 
near $\tan^2\theta=1$ (maximal mixing) all three regions meet
and form the large mixing angle
(LMA) solution. The small mixing angle solution (SMA) is the
``crossing point'' of the Gallium region and the Chlorine
region and lies
between $\Delta m^2\approx 10^{-6}$eV$^2$ and $10^{-5}$eV$^2$
at $\tan^2\theta\approx10^{-3}$.
The LOW solution is the large $\Delta m^2$ part of the
extensive
region(s) between $\approx10^{-7}$eV$^2$ and $10^{-9}$eV$^2$,
while the lower part of this region
is called the quasi-vacuum (quasi-VAC) solution.
Finally, the vacuum solutions (VAC) are found below $\approx10^{-10}$eV$^2$.
All regions have similar $\chi^2$
(though the LOW fits are slightly worse with
$\chi^2>\chi^2_{\mbox{\tiny min}}+3.1$).

\begin{table*}[p]
\caption{SK rate and uncertainty for eight energy bins.
The rates and statistical and systematic uncertainties (of the
spectrum shape) in the third column
are presented in units of SSM expectation. These systematic uncertainties
are assumed to be uncorrelated in energy.
The uncertainties in the fourth (uncertainty of the
$^8$B neutrino spectrum), fifth (uncertainty of the energy scale of 0.64\%)
and sixth column (uncertainty of the energy resolution of 2.5\%) are
fully correlated in energy (but uncorrelated with each other).
The combined uncertainty (last row) is
based on the sum of all bins of the zenith angle spectrum.
The combined rate has an additional systematic uncertainty of
$\sigma_{SK}=\plumin{2.9}{2.6}\%$~\cite{syserr}
(excluding $^8$B shape, energy scale and resolution),
which was added to the uncorrelated systematic uncertainties
of the spectrum shape.
\label{tab:zenspec}}
\begin{center}
\footnotesize
\begin{tabular}{|cc|c|ccc|}
\hline
Bin & Range [MeV] & Rate$\pm$stat$\pm$syst [SSM] &
$^8$B Spectrum & E-Scale & E-Resol. \cr
\hline
1 &  5.0-5.5  & 0.467$\pm$0.040$\plumi{0.017}{0.014}$ &
$\plumi{0.04}{0.02}\%$ & $\plumi{0.09}{0.01}\%$ & $\plumi{0.23}{0.21}\%$ \cr
2 &  5.5-6.5  & 0.458$\pm$0.014$\pm{0.007}$ &
$\pm{0.1}\%$           & $\pm{0.2}\%$           & $\pm{0.2}\%$ \cr
3 &  6.5-8.0  & 0.473$\pm$0.008$\pm{0.006}$ &
$\pm{0.4}\%$           & $\pm{0.6}\%$           & $\pm{0.2}\%$ \cr
4 &  8.0-9.5  & 0.460$\pm$0.009$\pm{0.006}$ &
$\pm{0.9}\%$           & $\pm{1.3}\%$           & $\pm{0.1}\%$ \cr
5 &  9.5-11.5 & 0.463$\pm$0.010$\pm{0.006}$ &
$\plumi{1.7}{1.6}\%$   & $\plumi{2.5}{2.4}\%$   & $\pm{0.2}\%$ \cr
6 & 11.5-13.5 & 0.462$\pm$0.017$\pm{0.006}$ &
$\plumi{3.1}{2.7}\%$   & $\plumi{4.4}{4.1}\%$   & $\pm{1.1}\%$ \cr
7 & 13.5-16.0 & 0.567$\pm$0.039$\pm{0.008}$ &
$\plumi{5.1}{4.2}\%$   & $\plumi{7.0}{6.4}\%$   & $\pm{3.2}\%$ \cr
8 & 16.0-20.0 & 0.555$\pm$0.146$\pm{0.008}$ &
$\plumi{7.7}{5.6}\%$   & $\plumi{10.6}
{\hspace*{5pt}9.6}\%$ & $\plumi{8.4}{7.9}\%$   \cr
\hline
\multicolumn{2}{|c|}{Comb. 5.0-20.0} &
                0.465$\pm{0.005}\plumi{0.014}{0.012}$ &
$\plumi{1.2}{1.0}\%$ & $\plumi{1.7}{1.6}\%$ & $\pm{0.3}\%$\cr
\hline
\end{tabular}
\end{center}
\caption{Subdivision of bins 2--7 according to the solar zenith angle
$\theta_z$. The range of $\cos \theta_z$ is given for each bin:
$\cos \theta_z<0$ is `Day' and $\cos \theta_z>0$ is `Night'
(`Mantle' and `Core').
The rates are given in units of 0.001$\times$SSM.
Only statistical uncertainties are quoted. All systematic uncertainties
(see Table~\ref{tab:zenspec}) are assumed to be fully correlated in
zenith angle.}
\label{tab:zenspec2}
\begin{center}
\footnotesize
\begin{tabular}{|c|c|ccccc|c|}
\hline
       & Day   & \multicolumn{5}{|c|}{Mantle} & Core \cr
\hspace*{-5pt}Bin\hspace*{-5pt}    & 
-0.974--0 & 0.00--0.16 & 0.16--0.33 & 0.33--0.50 & 0.50--0.67 & 0.67--0.84 & 0.84--0.974 \cr
\hline
2 & 453$\pm$20  & 442$\pm$53  & 379$\pm$49  &
    472$\pm$45  & 522$\pm$45  & 503$\pm$49  & 426$\pm$52  \cr
3 & 474$\pm$12  & 530$\pm$34  & 506$\pm$30  &
    438$\pm$26  & 478$\pm$26  & 451$\pm$28  & 439$\pm$31  \cr
4 & 448$\pm$13  & 463$\pm$36  & 470$\pm$33  &
    462$\pm$29  & 509$\pm$29  & 461$\pm$32  & 451$\pm$35  \cr
5 & 453$\pm$15  & 449$\pm$40  & 502$\pm$38  &
    451$\pm$32  & 473$\pm$32  & 477$\pm$35  & 483$\pm$40  \cr
6 & 477$\pm$25  & 509$\pm$67  & 351$\pm$55  &
    391$\pm$49  & 498$\pm$53  & 434$\pm$56  & 521$\pm$64  \cr
7 & 511$\pm$54  & 570$\pm$150 & 831$\pm$167 &
    694$\pm$131 & 665$\pm$127 & 441$\pm$118 & 469$\pm$131 \cr
\hline
1-8&459.9$\pm$6.7&483$\pm$ 18 & 476$\pm$17 &
    451$\pm$15  & 496$\pm$ 15 & 467$\pm$16 & 456$\pm$17 \cr

\hline
\end{tabular}
\end{center}
\end{table*}

\begin{figure*}[hbt]
\includegraphics[height=7.8cm,clip]{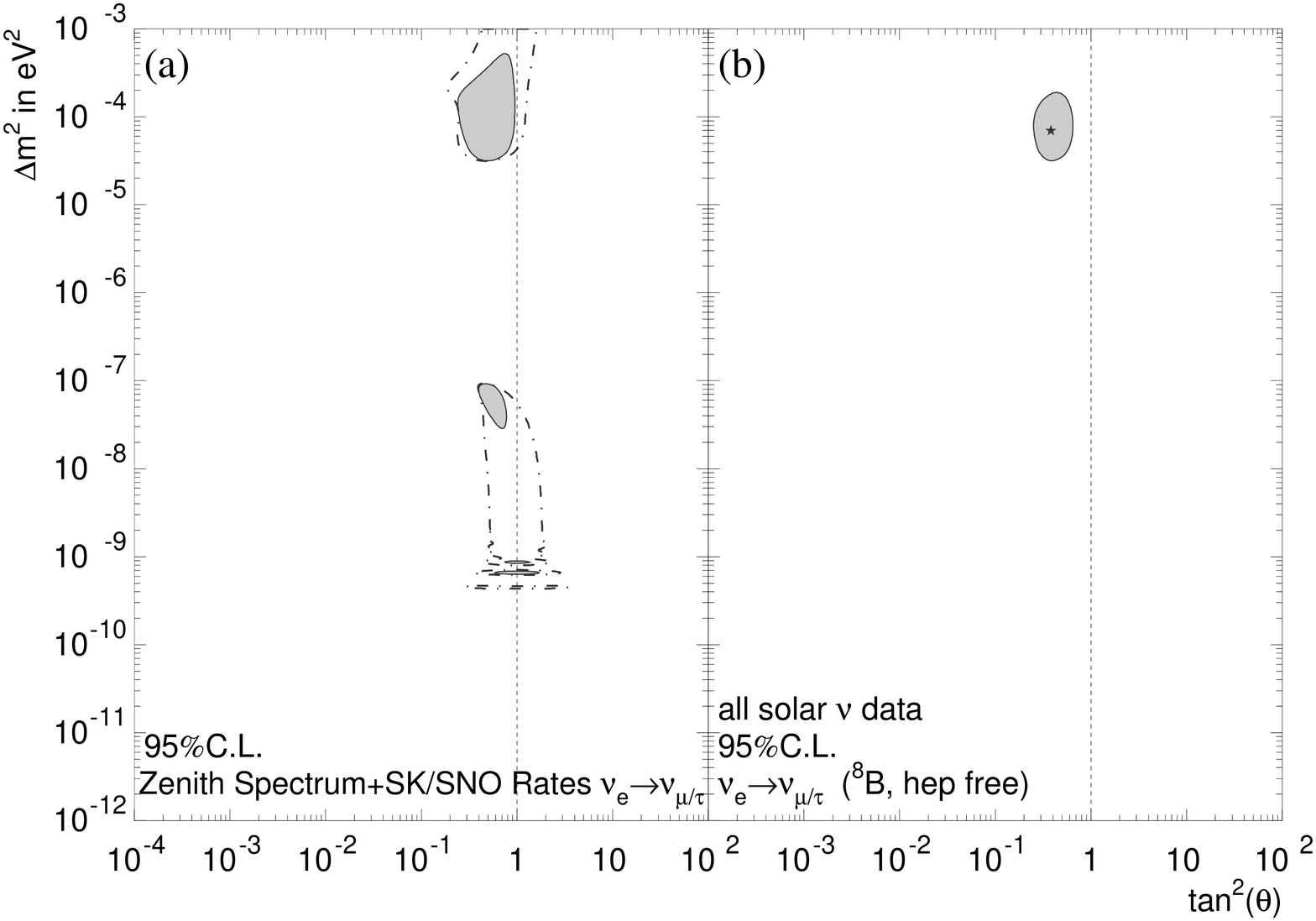}

\caption{
Solar neutrino oscillation parameter constraints
from combined fits using the SK zenith spectrum and
rate and other experimental rates.
(a) SK and SNO combined fit.
The dashed-dotted lines
indicate the allowed area if the SNO
neutral-current rate is removed from the fit.
(b) The combined fit to SK data 
and all rates favor the LMA solution.
The best fit point is indicated by the asterisk.
All fits are independent of the SSM $^8$B
flux and {\it hep} flux.
}
\label{fig:global}
\end{figure*}

SK data breaks the degeneracy in $\chi^2$ of these regions
by analyzing the
shape of the ``zenith angle spectrum''~\cite{osc} which
combines spectrum
and daily variation analyses.
Here we utilize the
zenith angle spectrum from the entire SK-I
data set collected between
May 31st, 1996 and July 15th, 2001
(1496 live days). 
The numerical results are listed in
Tables~\ref{tab:zenspec} and \ref{tab:zenspec2}.
There is no indication of spectral distortion, the
$\chi^2$ to an undistorted spectrum is 4.6 for 7
degrees of freedom (71\% C.L.).
No significant daily variation is found,
the day/night rate asymmetry is 
$A_{\mbox{\tiny DN}}=
\frac{\mbox{\tiny Day-Night}}{\mbox{\tiny 0.5(Day+Night)}}=
-0.021\pm$0.020(stat)$^{+0.013}_{-0.012}$(syst).
The gray excluded regions of 
Figure~\ref{fig:sk} (a) are the result of
the zenith angle spectrum fit.
The SMA and VAC regions are excluded since they require a
distorted $^8$B neutrino spectrum.
The lower part of the
LMA and the upper part of the LOW solution 
predict daily variations
and are therefore disfavored.
The excluded areas
are independent of the SSM neutrino fluxes. While the SK zenith
angle spectrum shape is best described by quasi-VAC
solutions, LMA fits have similar $\chi^2$.

Since there is no indication of a distortion of the SK zenith
spectrum, an analysis of the zenith spectrum shape alone can only yield
excluded regions. However, if the $^8$B flux is known,
the SK rate can be predicted. Then, the zenith spectrum shape
can be combined with the SK rate into a single measurement.
Figure~\ref{fig:sk}(b) shows the allowed regions
using SK rate and zenith angle spectrum in combination
with the $^8$B flux prediction and uncertainty of the SSM.
Unlike other experiments, whose allowed areas are shown in
Figure~\ref{fig:singleexp} (a), SK allows only
large mixing at 95\% C.L, and there are two
allowed regions: the LMA and quasi-VAC solution.

\section{Combination of Super-Kamiokande Data with Other Experiments}

A comparison between the SK rate
and SNO's charged-current rate~\cite{snonc}
yields another $^8$B flux constraint which is independent
of the SSM,
so a combination of the SK rate and zenith angle
spectrum with the SNO charged-current rate need not rely on any
neutrino flux prediction~\cite{venice}.
The SK rate
of $0.465\plumin{0.015}{0.013}\times$SSM
(see Table~\ref{tab:zenspec})
exceeds the SNO charged-current rate~\cite{snonc} of
$0.349\plumin{0.023}{0.022}\times$SSM,
by about $4.5\sigma$.
If this is interpreted as an appearance signal of other
active neutrino flavors,
these flavors contribute about 25\% to the SK rate and
70\% to the $^8$B flux
(the $\nu-e$ elastic scattering cross section for the other flavors
is six to seven times smaller than for $\nu_e$).
As a consequence of this appearance signal, the two allowed regions
(dashed-dotted lines) of
Figure~\ref{fig:global} (a)
result from
a combined fit: a LMA and a quasi-VAC solution. If the SNO
neutral-current rate is used as well, the quasi-VAC solutions
are disfavored as seen in
Figure~\ref{fig:global} (a)
(gray shaded areas).

To constrain the quasi-VAC region below the LOW
solution, it is necessary to
add the Gallium~\cite{gallex,sage}
($74.8\plumin{5.1}{5.0}$~SNU or
$0.584\plumin{0.040}{0.039}\times$SSM)
and Chlorine rates~\cite{cl} ($2.56\pm0.23$~SNU or
$0.337\pm0.030\times$SSM) to the fit.
Such a global fit is still not influenced by the $^8$B and
{\it hep} neutrino
(arising from $^3$He$+$p$\rightarrow ^4$He+e$^+$+$\nu_e$)
flux predictions of the SSM, which
suffer from the largest uncertainties. However, it relies
on the other SSM neutrino fluxes, in particular the $^7$Be
flux (10\% uncertainty) and the neutrino fluxes of the CNO
cycle ($\approx20\%$ uncertainty). Those fluxes contribute~\cite{ssm}
about 15\% ($^7$Be) and 6\% (CNO) to the Chlorine rate 
and 27\% ($^7$Be) and 7\% (CNO) to the Gallium rate in the SSM.
The allowed areas of this global fit shown in
Figure~\ref{fig:global} (b) looks quite different
when compared with the fit without SK data displayed 
in Figure~\ref{fig:singleexp} (b):
only the upper part of the LMA survives.

Table~\ref{tab:solution} compares the four smallest local minima 
of the $\chi^2$
describing the global fit. The best fit is located in the upper LMA area.
The $^8$B flux resulting from this fit 
is somewhat higher than expected by the SSM
($5.05\plumin{1.01}{0.81}\times10^6/$cm$^2\cdot$s)
but well within the uncertainty.
The {\it hep} flux 
is considerably higher than expected by the SSM
($9.3\times10^3/$cm$^2\cdot$s);
however, the uncertainty of this prediction is thought
to be very large.
The fit agrees with the SK zenith angle spectrum moderately well
and easily accommodates the Gallium, SK and SNO rates,
though the predicted
Chlorine rate is about $2\sigma$ too high. This worsens the otherwise very
good best fit $\chi^2$.

\begin{figure}[hbt]
\includegraphics[height=7.8cm,clip]{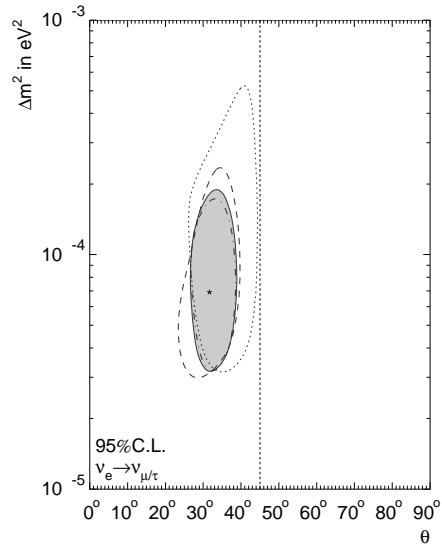}
\caption{
Magnified view of the LMA solution. Note the linear
scale of the x-axis. The 95\% C.L.
contours are obtained from the global fit with
(dashed, dotted and dashed-dotted lines) and
without (gray shaded area)
using the SSM $^8$B flux calculation.
The dashed-dotted lines include all experimental
information, while the
dashed (no SNO rates)
and dotted lines
(no radio-chemical rate measurements) do not.
}
\label{fig:lma}
\end{figure}

The quasi-VAC solution describes the SK zenith angle spectrum slightly
better than the LMA does, however, the rates do not fit well.
The resulting
best fit $^8$B flux falls $1.5\sigma$ short of the SSM prediction.
The surviving (at 99\% C.L.) LOW solution fits the rates about as poorly as
the quasi-VAC solution.
The SK zenith angle spectrum fits LOW worse than it does either LMA
or quasi-VAC.
The lack of zenith angle variation in the SK data reduces the $\Delta m^2$
(usually around $10^{-7}$eV$^2$) and worsens the LOW best fit which is already
under pressure from the rates.

At around the $3.5\sigma$ level a small mixing angle solution appears.
The mixing angle is about a factor of four smaller than the usual
SMA. For any given $^8$B flux, such a small mixing angle region can
be defined by the crossing of the Gallium and the Chlorine allowed
areas; those two rates therefore automatically fit very well.
In case of the classic SMA, the SK spectral data lack the predicted distortion.
As a consequence, the region gets shifted towards smaller mixing, and
even then the SK zenith angle spectrum considerably (about
3.5 $\sigma$) disfavors this fit. Due to the reduced mixing angle,
the predicted SNO charged-current rate is more than $2\sigma$ above the measurement.
Consequently, small mixing is excluded at the $3.5\sigma$ level.

\begin{table*}[htb]
\caption{Parameters for the best fit points. 
The probabilities given in the fourth row of numbers are
based on the difference of $\chi^2$
(with respect to the minimum).
The six rows below show the six independent
parts of the fit: the $\Delta\chi^2$ from a
fit to the shape of the SK zenith spectrum
and the five
interaction rates (deviation probabilities
are given in units of Gaussian standard deviation
$\sigma$ which contains only experimental uncertainties).
The last four rows show the values of the
minimized fit parameters.
The $^8$B and the {\it hep} fluxes are free,
while the $^8$B neutrino spectrum shift as well
as the SK energy scale and resolution shifts
are constrained within the systematic uncertainties.} 
\label{tab:solution}
\begin{center}
\footnotesize
\begin{tabular}{|l|cccc|}
\hline
Solution &
Large Mixing&\hspace*{-1mm}Quasi-Vacuum\hspace*{-1mm}&Low $\Delta m^2$& Small Mixing\cr
& Angle (LMA)     &   (Quasi-VAC)      &       (LOW)        & Angle (SMA) \cr
\hline
$\Delta m^2$                    &
6.9$\times10^{-5}$ & 6.68$\times10^{-10}$ & 7.2$\times10^{-8}$ & 6.6$\times10^{-6}$ \cr
$\tan^2\theta$                  &
0.38               & 1.5                  & 0.66               & 0.0012             \cr
\hline
$\chi^2$ (46 dof; $p_{\chi^2}$ [\%]) &
43.5  (57.7)       & 53.5 (20.9)          & 52.5 (23.7)        & 58.9  (9.6)        \cr
$\Delta\chi^2$(2 dof;$p_{\Delta\chi^2}$[\%]) &
 0.0 (100.0)       & 10.0  (0.7)          &  9.0  (1.1)        & 15.4  (0.05)
\hspace*{-7pt}        \cr
\hline
$\Delta\chi^2_{SK}$ ($p_{\Delta \chi^2}$[$\sigma$]) &
 3.0 ( $1.2\sigma$) &  1.0 ( $0.5\sigma$) & 5.6 ( $1.9\sigma$)  &15.1 ( $3.5 \sigma$) \cr
Ga Rate [SNU]                   & 
71.7 ($-0.6\sigma$)& 64.4  ($-2.1\sigma$) &64.9  ($-2.0\sigma$) &78.3  ($+0.7\sigma$)\cr
Cl Rate [SNU]                   &
2.99 ($+1.9\sigma$)&  3.09 ($+2.4\sigma$) & 3.03 ($+2.1\sigma$) & 2.43 ($-0.6\sigma$)\cr
SK Rate [\%SSM]                 &
46.1 ($-0.4\sigma$)& 45.0  ($-1.2\sigma$) &45.5  ($-0.8\sigma$) & 45.7 ($-0.6\sigma$)\cr
SNO CC [\%SSM]                  &
34.0 ($-0.4\sigma$)& 38.9  ($+1.8\sigma$) &37.7  ($+1.3\sigma$) & 40.0 ($+2.3\sigma$)\cr
SNO NC [\%SSM]                  &
106 ($+0.4\sigma$) &   76  ($-2.0\sigma$) &  87  ($-1.1\sigma$) &   89 ($-1.0\sigma$)\cr
\hline
$\phi_{^8B}$ [$10^6/($cm$^2$s)] &
5.33 ($+0.3\sigma$)& 3.83 ($-1.5\sigma$)  & 4.41 ($-0.8\sigma$)& 4.48 ($-0.7\sigma$)\cr
$\phi_{hep}$ [$10^3/($cm$^2$s)] &
36                 & 9                    & 23                 & 12                 \cr
$^8$B Spectrum Shape            &
$-0.3\sigma$       & $-0.3\sigma$         & $+0.0\sigma$       & $+0.9\sigma$       \cr
SK E-scale/resol.               &
$-0.5\sigma$/$-0.1\sigma$       & $-0.5\sigma$/ $0.2\sigma$ & 
$+0.0\sigma$/$-0.2\sigma$       & $+1.3\sigma$/$-0.3\sigma$ \cr
\hline
\end{tabular}
\end{center}
\end{table*}

Figure~\ref{fig:lma} shows a close up view of the LMA region.
The combined fits described above treat the $^8$B flux as a free parameter,
which is only constrained by the solar neutrino experiments.
If the SSM prediction of this flux is used (dashed-dotted line), the quasi-VAC, LOW and
small mixing solutions are further disfavored: the quasi-VAC solution 
is then excluded at 99.4\% C.L., LOW at 99.1\% C.L. and small mixing at 99.96\% C.L.
If the SNO measurements are removed from the fit (dashed lines), LMA is still favored.
Quasi-VAC solutions appear then at 95\% C.L, LOW solutions at 97\% C.L. and
small mixing solutions at 93\% C.L. If the radio-chemical measurements (Gallium
and Chlorine) are ignored (dotted lines, or see 
Figure~\ref{fig:global} a),
LOW solutions
appear at 92\% C.L., quasi-VAC at 84\% C.L. and small mixing solutions at 99.99\% C.L.

\section{Conclusion}
The combined fit to the charged-current interaction rates measured by
several experiments result in many allowed regions
of neutrino mixing and mass square difference:
LMA, SMA, LOW, quasi-VAC and VAC solutions.
The absence of spectral distortion and daily variations of the
SK solar neutrino interaction rate strongly constrains these regions
rejecting SMA, LOW
and VAC solutions at 95\% C.L., while leaving only the higher
mass square difference LMA and the quasi-VAC region.
When combined with the SK interaction rate and either
the SSM prediction of the $^8$B flux or the SNO rates,
large neutrino mixing is selected. 
When Gallium, Chlorine and SNO data are combined with SK data, only
the higher mass square difference LMA solutions remain at
98.9\% C.L. Thus the combined results of all solar neutrino
experiments can be used to determine a {\em unique} region
of oscillation parameters that explains the famous solar neutrino
problem.

\section*{Acknowledgments}
We gratefully acknowledge the cooperation of the Kamioka Mining and Smelting 
Company.
The Super-Kamiokande detector has been built and
operated from funding by the Japanese Ministry of Education, Culture,
Sports, Science and Technology, the U.S. Department of Energy, and the
U.S. National Science Foundation.


\begin{thebibliography}{99}
\newcounter{notes}
\addtocounter{notes}{1}
\bibitem[\fnsymbol{notes}]{Harvard}
Present address: Harvard University, Cambridge, MA 02138, USA
\addtocounter{notes}{1}
\bibitem[\fnsymbol{notes}]{EFI}
Present address: Enrico Fermi Institute, University of Chicago, 
Chicago, IL 60637, USA
\addtocounter{notes}{1}
\bibitem[\fnsymbol{notes}]{RIKEN}
Present address: The Institute of Physical and Chemical Reasearch (RIKEN), 
Wako, Saitama 351-0198, Japan
\addtocounter{notes}{1}
\bibitem[\fnsymbol{notes}]{Utah}
Present address: Department of Physics, University of Utah,
Salt Lake City, UT 84112, USA
\bibitem{cl}     B.T.~Cleveland et al.,    \Journal{\APJ}{496}{505}{1998}.
\bibitem{kam}    Y.~Fukuda et al.,         \Journal{\PRL}{77}{1683}{1996}.
\bibitem{gallex} E.~Bellotti, \Journal{\NPB\em(Proc. Suppl.)}{91}{44}{2001};\\
                 W.~Hampel et al.,         \Journal{\PLB}{388}{364}{1996};\\
                 P.~Anselmann et al.,      \Journal{\PLB}{342}{440}{1995}.
\bibitem{sage}   V.~Gavrin,   \Journal{\NPB\em(Proc. Suppl.)}{91}{36}{2001};\\
                 J.N.~Abdurashitov et al., \Journal{\PLB}{328}{234}{1994}.
\bibitem{ssm}    J.N.~Bahcall, M.H.~Pinsonneault, S.~Basu,
                                           \Journal{\APJ}{555}{990}{2001}. 
\bibitem{sk}     S.~Fukuda et al.,         \Journal{\PRL}{86}{5651}{2001}. 
\bibitem{sno}    Q.R.~Ahmad et al.         \Journal{\PRL}{87}{71301}{2001}.
\bibitem{skatm}  Y.~Fukuda et al.,         \Journal{\PRL}{81}{1562}{1998}.
\bibitem{chooz}  M.~Apollonio,             \Journal{\PLB}{466}{415}{1999}.
\bibitem{msw}    S.P.~Mikheyev and A.Y.~Smirnov, 
                                           \Journal{\SNP}{42}{913}{1985};
                 L.~Wolfenstein,           \Journal{\PRD}{17}{2369}{1978}.
\bibitem{osc}    S.~Fukuda et al.,         \Journal{\PRL}{86}{5656}{2001}.
\bibitem{snonc}  Q.R.~Ahmad et al.         {\em nucl-ex/0204008} (2002).
\bibitem{venice} M.~Smy in {\em Neutrino Oscillations in Venice},
                 ed.~M.~Baldo Ceolin, (Venezia, 2001) 35.
\addtocounter{notes}{1}
\bibitem[\fnsymbol{notes}]{syserr}
A detailed
discussion of this uncertainty will be published elsewhere.
\addtocounter{notes}{1}
\bibitem[\fnsymbol{notes}]{ap}
As we prepared this paper, SNO announced new results on its charged-
and neutral-current rates, which
assumes an undistorted $^8$B $\nu_e$ spectrum. In particular,
the neutral-current rate can be strongly influenced by such
spectral distortions. At present, the SNO measurement of this
spectrum cannot be used to extract neutral-current and charged-current
rates for distorted spectra, since the 
systematic uncertainties have not yet been given as a
function of energy. As a consequence, we do not use the neutral-current
rate in oscillation fits excluding SK data. In combined
fits including the SK recoil electron spectrum, the distortions
are small in the regions which are allowed by SK data.
\end{thebibliography}
\end{document}